\documentclass[useAMS,usenatbib,fleqn]{mn2e}
\usepackage{lscape,graphicx,natbib,amssymb,amsmath}
\usepackage{array}
\usepackage{url}
\usepackage{fixltx2e}
\usepackage[T1]{fontenc}
\usepackage{ae}
\usepackage{aecompl}
\newcolumntype{L}[1]{>{\raggedright\let\newline\\\arraybackslash\hspace{0pt}}m{#1}}
\newcolumntype{C}[1]{>{\centering\let\newline\\\arraybackslash\hspace{0pt}}m{#1}}
\newcolumntype{R}[1]{>{\raggedleft\let\newline\\\arraybackslash\hspace{0pt}}m{#1}}
\usepackage{array}
\title[A new analysis of fine-structure constant measurements]
  {A new analysis of fine-structure constant measurements and modelling errors from quasar absorption lines}
\author[M. R. Wilczynska et al.]
  {Michael.~R.~Wilczynska,$^1$\thanks{E-mail: mikew@phys.unsw.edu.au}
  John~K.~Webb,$^1$\thanks{E-mail: jkw@phys.unsw.edu.au}
  Julian~A.~King,$^1$
  \newauthor
  Michael~T.~Murphy,$^2$
  Matthew~B.~Bainbridge,$^1$
  and Victor~V.~Flambaum,$^1$  \\
  $^1$School of Physics, University of New South Wales, Sydney, NSW 2052, Australia\\
  $^2$Centre for Astrophysics and Supercomputing, Swinburne University of Technology, Victoria 3122, Australia\\}
\date{Released 2014 Xxxxx XX}

\pagerange{\pageref{firstpage}--\pageref{lastpage}} \pubyear{2014}

\def\LaTeX{L\kern-.36em\raise.3ex\hbox{a}\kern-.15em
    T\kern-.1667em\lower.7ex\hbox{E}\kern-.125emX}

\begin{document}

\label{firstpage}

\maketitle

\begin{abstract}

We present an analysis of 23 absorption systems along the lines of
sight towards 18 quasars in the redshift range of $0.4 \leq z_{abs} \leq
2.3$ observed on the Very Large Telescope (VLT) using the Ultraviolet
and Visual Echelle Spectrograph (UVES). Considering  both statistical
and systematic error contributions we find a robust estimate of the
weighted mean deviation of the fine-structure constant from its current,
laboratory value of
$\Delta\alpha/\alpha=\left(0.22\pm0.23\right)\times10^{-5}$, consistent
with the dipole variation reported in \cite{WKM+10} and \cite{Kin12}.

This paper also examines modelling methodologies and systematic effects.
In particular we focus on the consequences of fitting quasar absorption
systems with too few absorbing components and of selectively fitting
only the stronger components in an absorption
complex. We show that using insufficient continuum regions around an
absorption complex causes a significant increase in the scatter of a sample
of $\Delta\alpha/\alpha$ measurements, thus unnecessarily reducing the overall
precision.  We further show that
fitting absorption systems with too few velocity components also results
in a significant increase in the scatter of $\Delta\alpha/\alpha$
measurements, and in addition causes $\Delta\alpha/\alpha$ error estimates to be
systematically underestimated. These results thus identify some of the 
potential pitfalls in analysis techniques and provide a guide for future
analyses.

\end{abstract}


\section{Introduction}

Modern physical theories utilise a set of fundamental constants. These constants are defined as dimensionless ratios of physical quantities and have no dependence on other measured parameters. As such, it is important to know the numerical values of these constants to high accuracy. \cite{Wey19}, \cite{Edd31} and \cite{Dir37} were amongst the first to investigate the possibility that these fundamental constants may vary with time. More recently, attempts at deriving unifying theories of nature have revived interest in varying constants, as some of these theories predict the variation of fundamental constants. For a comprehensive review of both theory and observations see \cite{Uza11}. This work focuses on the fine-structure constant  $\alpha \equiv e^2/4 \pi \epsilon_0 \hbar c$ and its possible variation:

\begin{equation}
\Delta\alpha/\alpha\equiv  \frac{\alpha_{z}-\alpha_{0}}{\alpha_{0}},
\end{equation}

\noindent comparing the laboratory value, $\alpha_{0}$, with the value at some redshift, $\alpha_{z}$.

Quasar absorption line systems provide important probes
of the possible variation of the fine-structure constant over
cosmological time-scales and distances. These analyses exploit the absorption features caused by intervening gas clouds along the lines of sight to background quasars. Possible evidence for a varying $\alpha$ has surfaced in recent years. These studies employed the `Many-Multiplet' (MM) method \citep{DFW99,WFC+99}, beginning with 30 absorption systems observed on the Keck telescope \citep{WFC+99} and later extended to 140 absorption systems by \cite{MWF+03} and \cite{MFW+04}. The next large study used observations from the Very Large Telescope (VLT) consisting of 154 measurements of $\alpha$ \citep{Kin12}. Combining the results of the Keck sample with those of the VLT sample, \cite{WKM+10} and \cite{Kin12} found that the data indicated a variation in $\alpha$ across the sky that could be consistently modelled with a spatial dipole. If correct, clearly the result is of fundamental significance and hence must be critically tested and the associated systematic effects comprehensively understood.

The quasar absorption system sample investigated in this work has been previously studied, with a discussion 
arising in the literature about the validity of the analysis \citep{CSP+04,MWF07,SCP+07,MWF08}. 
We present an independent analysis of this sample of quasar absorption systems.

This work will examine, in greater detail
compared to previous work, methodologies associated with absorption line
modelling techniques. To explore the systematic effects of differing fitting methods
on $\Delta\alpha/\alpha$ measurements we model the 23
absorption systems in three different ways. We also discuss how
different analysis strategies can significantly influence uncertainties
in measurements of a fractional change in $\alpha$.

We choose to work with this sample of absorbers, originally analysed by \citet{CSP+04} (hereafter C04), for a variety of reasons: the absorption systems within this sample have relatively simple velocity structures; the sample is large enough such that meaningful statistical analysis may be applied to gain insight into the effects of differing fitting methodologies; and finally, this sample has been previously studied thereby allowing for a comparison of previous results with the independent analysis of this work.

A potentially important source of systematic errors in modelling
quasar absorption systems to derive estimates of space-time variation
of $\alpha$ concerns possible spectral wavelength distortions.  
\citet{GWW10} first discovered small-scale, i.e. intra-order, wavelength calibration problems.  
\citet{MLM08} first searched for possible wavelength distortions across echelle
orders by comparisons with echelle calibrations from asteroids measurements.
\citet{RWS2013} subsequently discovered large-scale, approximately linear 
distortion effects and a more detailed study was carried out by \cite{EMW14}. \cite{EMW14} compared observations from different telescopes of the same quasar and found long range distortions of the quasar spectra's wavelength scales.  Recently \citet{MW15} applied a simple
distortion model to quasar simulations and concluded that long range wavelength distortions are capable of significantly weakening the evidence for variations in $\alpha$ from quasar absorption lines.  We note that
the \citet{MW15} model effectively assumes a single-wavelength setting,
inappropriate for most quasar observations.  This assumption is critically examined 
in a parallel study \citep{DW15} who reach different conclusions and we therefore do not include distortion models presented in \citet{MW15}.

\subsection{Overview}

In Section 2, we describe the spectroscopic and atomic data used in our analysis. Section 3 describes the profile analysis methods used in this work. Section 4 describes the Least Trimmed Squares (LTS) Method, which is used to estimate a random, systematic error we incorporate into our final results. Section 5 presents our methodology for and our independent analysis of the dataset. Section 6 describes two additional fitting methodologies we apply to the entire dataset and the quantitative effects of applying differing Voigt profile fitting methodologies on measurements of $\Delta\alpha/\alpha$.

\section{Spectroscopic and atomic data}
\label{sec:Spectroscopic and atomic data}

The raw spectroscopic data we use in this work is the same as presented in \citet{CSP+04}. 
However, we have independently reduced the raw spectra to produce the 1-dimensional calibrated 
spectra as described below. We have not included any additional data other than that used 
in \citet{CSP+04}. All spectral data are taken from the VLT/UVES archive. Raw spectra were
reduced to 1-dimensional format and calibrated using the \textsc{midas}
pipeline, provided by ESO. We have used the updated echelle spectrograph ThAr line list detailed 
in \citet{MTW+07}. We have followed the same procedures for these and all other
aspects of data reduction and calibration, as described in \citet{Kin12}.

The \textsc{midas} extraction routine incorrectly estimates the spectral
errors associated with the flux of data points in the base of saturated
absorption lines. When individual exposures are combined
\textsc{uves\_popler} \footnote{\url{http://astronomy.swin.edu.au/~mmurphy/UVES_popler/}}
provides a check on the concordance between the
different exposures. \textsc{uves\_popler} calculates the value of
$\chi_{\nu}^{2}$ for each combined flux pixel by
considering the dispersion of the contributing pixels from each exposure 
about their weighted mean. The bases of saturated lines
typically produce a $\chi_{\nu}^{2}\geq{2}$, suggesting that
\textsc{midas} spectral error arrays underestimate the statistical uncertainty in those
regions. The problem is more noticeable towards the blue end of spectra.
The effect of this is to underestimate the statistical uncertainty on
$\Delta\alpha/\alpha$ derived from these data points. To rectify this
problem we adjust the error arrays to account for the inconsistency
present in our spectra (see Section 2.1 of \citet{Kin12}).

\begin{figure*}
 \vspace{0.5cm}
\includegraphics[width=162mm,angle=0]{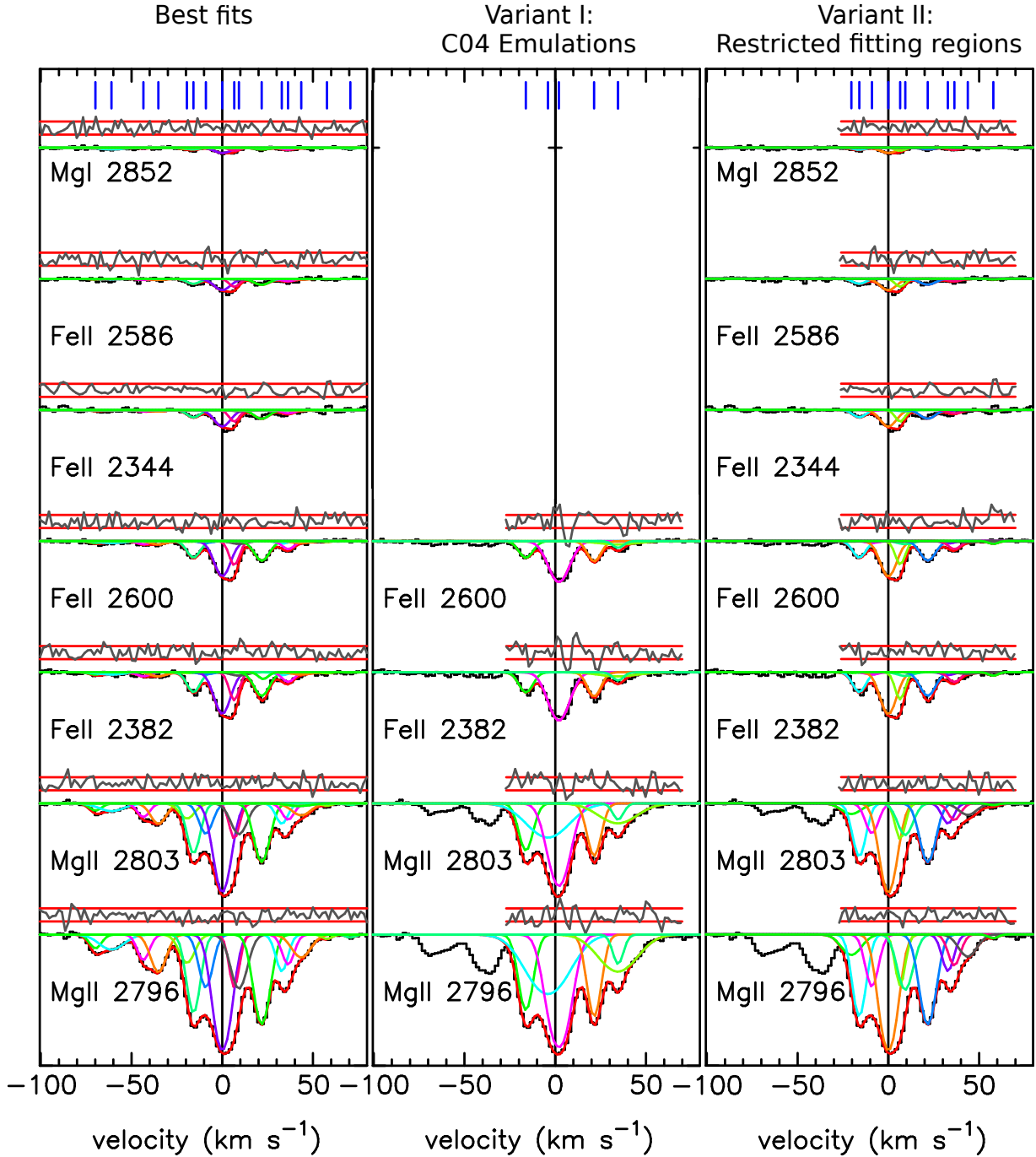}
  \caption{All three columns show $z_{abs}=0.942$ absorption
system along the line of sight to quasar J222006-280323. The first column shows
our best fits to the data. The second column shows our C04 emulations 
(Variant I). The third column shows our 
best fits but with restricted fitting regions (Variant II). The models to the data are shown 
as red lines. The individual components are all plotted in different colours. Normalised residuals 
({[}model\,-\,data{]}/error) are shown in grey between the horizontal red $\pm1\sigma$ error 
lines. Blue tick marks indicate individual velocity components. The best fit give a result of $\Delta\alpha/\alpha=(0.686\pm1.130)\times10^{-5}$, with a $\chi_{\nu}^{2}=0.947$. Variant I 
gives a result of $\Delta\alpha/\alpha=(-1.400\pm0.508)\times10^{-5}$, with a 
$\chi_{\nu}^{2}=2.480$. Variant II gives a result of $\Delta\alpha/\alpha=(0.068\pm1.010)\times10^{-5}$, 
with a $\chi_{\nu}^{2}=1.009$. The normalised residuals illustrate the limits of the fitting regions. 
Two contaminants are present in our best fit. One weak, but warranted by AICc, contaminant in Mg\,\textsc{ii} $\lambda$2803 at $v$ $\approx$ 35$\>$km$\;$s\textsuperscript{-1} and another in Fe\,\textsc{ii} $\lambda$2383
at $v$ $\approx$ 18$\>$km$\;$s\textsuperscript{-1}. Note the difference in fitting regions
when comparing the Variant I system to that of our best fit. The fitting
regions for Variant I extend only over the strongest absorption features, excluding absorption in the system that is clearly present. No continuum regions have been included for Variant I. 
Parameters describing the varying continuum and varying 
zero-level are not shown, but fitting parameters for all three models are provided in the supplementary material.
\label{m1m2m3}}
\end{figure*}

\begin{table}
 \centering
  \caption{Selected atomic data. Column 1 shows the common names of transitions. Wavelength ID's are provided in column 2. Actual wavelengths used in this work are compiled from isotopic structure, where available, and are taken from Table B1 of \citet{Kin12} (also see Section \ref{sec:Profile analysis methods}). \textit{q}-coefficients, which determine the sensitivity of atomic transitions to a change in $\alpha$ \citep{DFW99,MWF03}, are given in column 3. The letters in column 4 offer a simple shorthand for labeling transitions used to fit absorption systems. Our codes differ from those in \citet{Kin12} for clarity. \citet{Kin12} provide a thorough compilation of atomic data used in this work.\label{tab:atomdata}}
  \label{symbols}
  \begin{tabular}{@{}llcl}
  \hline
  Ion & \multicolumn{1}{c}{\textit{$\lambda_{ID}$}} & \multicolumn{1}{l}{\textit{q}[cm]$^{-1}$} & \multicolumn{1}{c}{Code} \\
  \hline
  Fe\,\textsc{ii}  & $1608$ & $-1300(300)$ & $a_{1}$ \\
                  & $2260$ & $1435(150)$  & $a_{2}$ \\
                  & $2344$ & $1210(150)$  & $a_{3}$ \\
                  & $2374$ & $1590(150)$  & $a_{4}$ \\
                  & $2382$ & $1460(150)$  & $a_{5}$ \\
                  & $2586$ & $1490(150)$  & $a_{6}$ \\
                  & $2600$ & $1330(150)$  & $a_{7}$ \\
  Mg\,\textsc{i}   & $2852$ & $86(10)$     & $b_{1}$ \\
  Mg\,\textsc{ii}  & $2796$ & $211(10)$    & $c_{1}$ \\
                  & $2803$ & $120(2)$     & $c_{2}$ \\
  Al\,\textsc{ii}  & $1670$ & $270(30)$    & $d_{1}$ \\
  Al\,\textsc{iii} & $1854$ & $464(30)$    & $e_{1}$ \\
                  & $1862$ & $216(30)$    & $e_{2}$ \\
  Si\,\textsc{ii}  & $1526$ & $50(30)$     & $f_{1}$ \\
                  & $1808$ & $520(30)$    & $f_{2}$ \\
  Mn\,\textsc{ii}  & $2576$ & $1420(150)$  & $g_{1}$ \\
                  & $2594$ & $1148(150)$  & $g_{2}$ \\
                  & $2606$ & $986(150)$   & $g_{3}$ \\
  Ni\,\textsc{ii}  & $1709$ & $-20(250)$   & $h_{1}$ \\
                  & $1741$ & $-1400(250)$ & $h_{2}$ \\
  Cr\,\textsc{ii}  & $2056$ & $-1110(150)$ & $i_{1}$ \\
                  & $2026$ & $-1280(250)$ & $i_{2}$ \\
                  & $2066$ & $-1360(250)$ & $i_{3}$ \\
  Zn\,\textsc{ii}  & $2026$ & $2479(25)$   & $j_{1}$ \\
                  & $2062$ & $1584(25)$   & $j_{2}$ \\
\hline
\end{tabular}
\end{table}

\textit{q}-coefficients of atomic transitions, which determine the sensitivity 
to $\alpha$ variation for a given atomic transition, wavelength ID's and ion codes used in 
this work are presented in Table \ref{tab:atomdata}. The atomic data in Table \ref{tab:atomdata} are taken from Table B1 of \citet{Kin12}, although we use different ion codes. We make 
use of isotopic structure where available (relevant for Si\,\textsc{ii}, Mg\,\textsc{i}, 
Mg\,\textsc{ii}, Zn\,\textsc{ii} in this work). We note that although there may be hints of a departure from terrestrial isotopic relative abundances for Mg at high redshift, we do not regard existing measurements as definitive.  For this reason, and because the analysis in this paper focuses on systematics associated with two particular issues (under-fitting and restricted spectral fitting regions), we have carried out the modelling in this paper using terrestrial relative abundances throughout.

\begin{table*}
 \centering
 \begin{minipage}{130mm}
  \caption{Transitions used in all fits contributing to the determination of $\Delta\alpha/\alpha$.
Columns 3 and 4 gives the quasar emission redshift and absorption
redshift respectively. The key for transition labels in column 5 and
6 is given in Table \ref{tab:transitions}. The $b$-parameters and redshifts for each component of 
an absorption system were tied in the fitting process for all species as described in Section \ref{sec:thermturb}, with the exception of Al\,\textsc{iii}. Al\,\textsc{iii} was fitted simultaneously, where available, 
but it's $b$-parameters and redshifts were fitted independantly. \label{tab:transitions}}
  \begin{tabular}{@{}llllcc@{}}
  \hline
  \multicolumn{2}{c}{Quasar Name}  &    &   \multicolumn{3}{c}{Transitions}  \\
  \multicolumn{1}{c}{J2000} & \multicolumn{1}{c}{B1950} & $\textit{z}_{em}$ & $\textit{z}_{abs}$ & Best fits and Variant II & Variant I \\
   \hline
 J134427-103541 & HE 1341-1020 & 2.35 & 0.873 & $a_{3}a_{4}a_{5}b_{1}c_{1}c_{2}$ & $a_{3}a_{5}a_{6}b_{1}c_{1}c_{2}$ \\
 J134427-103541 & HE 1341-1020 & 2.35 & 1.278 & $a_{2}a_{3}a_{4}a_{7}b_{1}c_{1}c_{2}g_{1}g_{3}$ & $a_{3}a_{4}a_{5}a_{7}c_{1}c_{2}$ \\
 J134427-103541 & HE 1341-1020 & 2.35 & 1.915 & $a_{1}a_{3}a_{4}a_{5}a_{7}f_{1}$ & $a_{1}a_{3}a_{4}a_{5}a_{7}f_{1}f_{2}$ \\
 J012417-374423 & Q 0122-380   & 1.91 & 0.822 & $a_{3}a_{5}a_{6}a_{7}b_{1}c_{1}c_{2}$ & $a_{3}a_{5}a_{6}a_{7}b_{1}c_{1}c_{2}$ \\
 J012417-374423 & Q 0122-380   & 1.91 & 0.859 & $a_{3}a_{5}a_{6}a_{7}b_{1}c_{1}c_{2}$ & $a_{5}a_{6}a_{7}b_{1}c_{1}c_{2}$ \\
 J012417-374423 & Q 0122-380   & 1.91 & 1.243 & $a_{2}a_{3}a_{4}a_{5}b_{1}c_{1}e_{1}e_{2}$ & $a_{3}a_{4}a_{5}c_{1}c_{2}$ \\
 J024008-230915 & PKS 0273-23  & 2.23 & 1.635 & $a_{1}a_{3}a_{6}b_{1}c_{1}c_{2}e_{1}e_{2}f_{1}$ & $a_{3}a_{6}a_{7}c_{1}c_{2}f_{1}$ \\
 J024008-230915 & PKS 0273-23  & 2.23 & 1.637 & $a_{1}a_{3}a_{6}a_{7}b_{1}c_{1}c_{2}$ & $a_{1}a_{3}a_{6}a_{7}c_{1}c_{2}$ \\
 J024008-230915 & PKS 0273-23  & 2.23 & 1.657 & $a_{3}a_{5}b_{1}c_{2}d_{1}f_{1}$ & $a_{3}a_{5}d_{1}f_{1}$ \\
 J000344-232355 & HE 0001-2300 & 2.28 & 0.452 & $a_{3}a_{5}a_{7}b_{1}c_{1}c_{2}$ & $a_{4}a_{5}a_{7}b_{1}c_{1}c_{2}$ \\
 J000344-232355 & HE 0001-2300 & 2.28 & 2.185 & - & $a_{3}a_{4}a_{5}c_{1}c_{2}d_{1}f_{1}$ \\
 J000344-232355 & HE 0001-2300 & 2.28 & 2.187 & - & $a_{3}a_{4}a_{6}c_{2}d_{1}f_{1}$ \\
 J011143-350300 & Q 0109-3518  & 2.40 & 1.182 & $a_{3}a_{5}a_{7}b_{1}c_{1}c_{2}$ & $a_{3}a_{5}a_{7}b_{1}c_{1}c_{2}$ \\
 J011143-350300 & Q 0109-3518  & 2.40 & 1.350 & $a_{3}a_{5}a_{6}a_{7}b_{1}c_{1}c_{2}$ & $a_{3}a_{4}a_{5}a_{6}a_{7}b_{1}d_{1}f_{1}$ \\
 J222006-280323 & HE 2217-2818 & 2.41 & 0.942 & $a_{3}a_{5}a_{6}a_{7}b_{1}c_{1}c_{2}$ & $a_{5}a_{7}c_{1}c_{2}$ \\
 J222006-280323 & HE 2217-2818 & 2.41 & 1.556 & $a_{3}a_{5}a_{6}a_{7}c_{1}d_{1}$ & $a_{3}a_{5}a_{6}a_{7}c_{1}c_{2}f_{1}$ \\
 J135038-251216 & HE 1347-2457 & 1.44 & 1.439 & $a_{1}a_{2}a_{3}c_{1}c_{2}f_{2}g_{2}i_{1}i_{2}i_{3}j_{1}j_{2}$ & $a_{3}a_{6}f_{1}f_{2}$ \\
 J045523-421617 & Q 0453-423   & 2.66 & 0.908 & $a_{3}a_{7}b_{1}c_{2}$ & $a_{3}a_{5}a_{7}c_{1}c_{2}$ \\
 J045523-421617 & Q 0453-423   & 2.66 & 1.858 & $a_{3}a_{5}a_{7}c_{1}d_{1}$ & $a_{3}a_{5}a_{7}c_{1}c_{2}$ \\
 J000448-415728 & Q 0002-422   & 2.76 & 1.542 & $a_{3}a_{4}a_{5}a_{7}c_{2}$ & $a_{3}a_{5}a_{6}a_{7}c_{1}c_{2}$ \\
 J000448-415728 & Q 0002-422   & 2.76 & 2.168 & $a_{1}a_{3}c_{1}c_{2}d_{1}f_{1}$ & $a_{1}a_{3}a_{5}a_{6}a_{7}d_{1}f_{1}$ \\
 J000448-415728 & Q 0002-422   & 2.76 & 2.302 & $a_{1}f_{1}h_{1}h_{2}$ & $a_{1}a_{3}a_{5}d_{1}f_{1}$ \\
 J212912-153841 & PKS 2126-150 & 3.29 & 2.022 & - & $a_{3}a_{5}a_{6}c_{2}f_{1}$ \\
   \hline
  \end{tabular}
 \end{minipage}
\end{table*}

\section{Profile analysis methods}
\label{sec:Profile analysis methods}

In Sections 5 and 6 we use \textsc{vpfit}\footnote{\url{http://www.ast.cam.ac.uk/~rfc/vpfit.html}} version 10.0 as our $\chi^{2}$ minimization algorithm. \textsc{vpfit} has been shown to be robust in its calculation of statistical uncertainties using Markov Chain Monte Carlo methods \citep{KMW+09}. The instrumental settings for the data used in this
analysis, assuming a Gaussian instrumental profile, correspond to a FWHM of 6$\>$km$\;$s\textsuperscript{-1}.

Estimates of $\alpha$ are derived using the Many Multiplet method \citep{DFW99,WFC+99}, modelling all transitions simultaneously. Table \ref{tab:transitions} provides a key for identifying which transitions are used in each absorption system and model.
$\Delta\alpha/\alpha$ is explicitly included as an additional free parameter which is varied along with all other parameters to minimize $\chi^{2}$.

All profile fits used in this work are made available in the supplementary material.

\section{The Least Trimmed Squares (LTS) method}
\label{sec:The Least Trimmed Squares (LTS) method}

Following \citet{Kin12} (Section 3.5.3) we use a LTS procedure to explore the
statistical properties of our whole dataset. That is, once we have measured $\Delta\alpha/\alpha$ in each absorber for each model we apply the LTS algorithm to the set of 23 $\Delta\alpha/\alpha$ measurements for each model. The purpose of applying the LTS method is twofold: to check whether there exist any outliers in the datasets; and, to determine an appropriate systematic error term to be added to each measurement of $\Delta\alpha/\alpha$ in quadrature. \citet{Kin12} utilised a fast-LTS algorithm \citep{RD06}, which computes orders of magnitude faster for large data sets compared to an LTS algorithm that computes the exact LTS procedure. We use an exact 
LTS algorithm as our dataset is small enough to accommodate an exhaustive 
search. Applying the LTS method removes the impact of any anomalous data 
points and is likely to provide a more robust estimate  of 
$\Delta\alpha/\alpha$.  Individual error estimates on $\Delta\alpha/\alpha$ 
returned by \textsc{vpfit} only account for 
the statistical properties of the spectral noise and do not account
for unknown systematics.  For that reason, we introduce
an additional error component added in quadrature to the
\textsc{vpfit} error term such that $\sigma_{total}^{2}=\sigma_{stat}^{2}+\sigma_{rand}^{2}$. 
We use the LTS method to allow for the inclusion, and calculation, of $\sigma_{rand}$.

Instead of fitting $n$ data points, the LTS method fits only $k=(n+p+1)/2$ points (where $p$ is the number of free parameters) using standard least-squares minimization and then searches for the combination of $k$ data points and fitted model that yields the lowest sum of squared residuals. The method only fits the inner $k/n$ fraction of the distribution of the residuals. Where up to $k-n$ outliers exist, they will be ignored in the calculation of $\sigma_{rand}$ by this method provided they are in the excluded fraction \citep{Rou84}.

To summarise, the steps are
as follows:
\begin{enumerate}

\item We select all possible subsets of $n$ from $k$ observations.
\item For each combination we compute the weighted mean value of $\Delta\alpha/\alpha$,
varying $\sigma_{rand}$ in the quadrature error term such that the observed
$\chi_{\nu}^{2}$ is equal to the expected value, given by Eq.17 in \citet{Kin12}.
$\sigma_{rand}$ is then determined in a more robust manner, without undue 
influence from anomalously deviant points, if any.  
\item Finally the $\sigma_{rand}$ deemed to be applicable to the whole
sample is selected as the smallest value out of all possible combinations. Selecting the smallest value of $\sigma_{rand}$ ensures that it's value is not overestimated by high scatter points and that any possible outliers are not `masked' by an artificially inflated $\sigma_{rand}$.
In this way we into take into account, in a robust way, any possible
systematic contribution to the overall error on $\Delta\alpha/\alpha$.
\item After applying $\sigma_{rand}$ to the entire dataset we search for any points 
with $|r_{i}|=|(\Delta\alpha/\alpha - \textrm{model\enspace prediction})/\sigma_{tot}|\geq3$ as outliers. However, none were found in this analysis.
\end{enumerate}

Note that for point (ii) above the LTS method increases $\sigma_{rand}$ sufficiently enough to reduce $\chi_{\nu}^{2}$ to an expected value and not unity. The expected value of $\chi_{\nu}^{2}$ is lower than unity due to the fact that the sample in question is a trimmed subset ($k/n$) of the original sample. In this situation it is possible for a sample with a weighted mean giving $\chi_{\nu}^{2}$ of unity to still require a systematic, non-zero, error term to be applied to derive a more robust estimate of the error term of that weighted mean (see Table \ref{table:weighted means} and its footnote).

\section{Profile fitting analysis: Best fits}
\label{sec:Independent analysis of the dataset (Model I)}

These absorption system models are our best 
Voigt profile fits to the data. The velocity structures for each absorption system in this model have been determined completely independently of those shown in C04.

\subsection{Fitting regions}
\label{sec:Fitting regions}

We choose fitting regions such that the flux recovers to the continuum and we 
include continuum regions either side of the absorption. Including ample continuum 
regions is important because: (i) if the model at the edge of the truncated region is significantly different from unity the convolution algorithm of \textsc{vpfit} will not return a reliable flux estimate for the pixels near the truncated edge, (ii) it prevents the clipping of the wings of absorption 
features, assuring \textit{all} parts of the absorption cloud are modelled, and (iii) we can 
allow for a varying continuum as an additional free parameter in the fit, if necessary. 

\subsection{Number of Voigt profile components used in a fit}
\label{sec:The Akaike Information Criterion (AIC)}

We follow the fitting procedure described in \citet{Kin12} to choose the optimal number of components in the 
fit. We use the Akaike Information Criterion (AIC) corrected for finite sample sizes \citep{Sig78}, defined as

\begin{equation}
AICc=\chi^{2}+2p+\frac{2p(p+1)}{(n-p-1)},
\end{equation}

\noindent where \textit{p} is the number of free parameters and \textit{n} is the number of data points included in a fit 
 \citep{Aka74}. We explore a range of models with different numbers of Voigt profile components, the adopted final model being the one with the smallest AICc. We only add a Voigt profile component to a fit if the addition of the component lowers the numerical value of the AICc. Employed this way, the AICc simply provides an additional check against fitting too many Voigt profile components to an absorption system.
 
 \begin{table*}
 \centering
 \begin{minipage}{167mm}
 \caption{\label{tab:results}Results on $\Delta\alpha/\alpha$ for many-multiplet absorption system fits. Errors
given are $1\sigma$ $(\sigma_{stat})$ and are taken directly from
the covariance matrix diagonal terms (i.e. not corrected for random
systematic effects). Column 2 lists the emission redshift, $z_{em}$,
of the quasar. Column 3 lists the absorption system redshift, $z_{abs}$.
Columns 4 and 5 list the $\chi_{\nu}^{2}$, $\Delta\alpha/\alpha$
and associated error in units of $10^{-5}$ for our best
fits to the absorption systems. Columns 6 and 7 list the same results for
Variant I (C04 emulations) and columns 8 and 9 list the
results for Variant II (restricted spectral fitting regions). We can predict 
the expected $\Delta\alpha/\alpha$ values for a given sight-line and model of 
$\alpha$ variation. Column 10 shows the predicted values of $\Delta\alpha/\alpha$ 
according to the \protect\cite{Kin12} dipole model $\Delta\alpha/\alpha=Acos(\Theta)+m$. 
Absorption systems marked with an asterisk were not included in the analysis for 
our best fit and Variant II models (see Section \ref{sec:Systems rejected from analysis}).}
  \begin{tabular}{@{}llllrlrlrc@{}}
  \hline
   Quasar Name      &  $z_{em}$  & $z_{abs}$   &      \multicolumn{2}{c}{Best fits} 
   & \multicolumn{2}{c}{Variant I}
   & \multicolumn{2}{c}{Variant II} & \multicolumn{1}{c}{Dipole Prediction} \\
   J2000  &  &  & $\chi_{v}^{2}$
     & $\Delta\alpha/\alpha[10^{-5}]$ & $\chi_{v}^{2}$ & $\Delta\alpha/\alpha[10^{-5}]$ 
     & $\chi_{v}^{2}$ & $\Delta\alpha/\alpha[10^{-5}]$ & $\Delta\alpha/\alpha[10^{-5}]$\\
 \hline
 J134427-103541 & 2.35 & 0.873 & 0.803 &  2.990$\pm$1.770 & 1.733 &  2.220$\pm$0.544 & 0.929 &  4.500$\pm$1.780 &  $\:$0.251 \\
 J134427-103541 & 2.35 & 1.278 & 0.827 &  0.337$\pm$1.130 & 3.797 &  3.900$\pm$1.140 & 1.187 &  0.438$\pm$1.130 &  $\:$0.251 \\
 J134427-103541 & 2.35 & 1.915 & 0.734 &  0.459$\pm$0.968 & 1.133 &  0.217$\pm$0.670 & 0.747 &  0.191$\pm$0.917 &  $\:$0.251 \\
 J012417-374423 & 1.91 & 0.822 & 0.794 &  0.923$\pm$1.090 & 1.035 &  1.220$\pm$1.100 & 1.047 &  1.210$\pm$1.100 &  $\:$0.145 \\
 J012417-374423 & 1.91 & 0.859 & 0.878 &  2.160$\pm$2.070 & 2.157 & -1.850$\pm$0.936 & 1.864 &  2.000$\pm$1.970 &  $\:$0.145 \\
 J012417-374423 & 1.91 & 1.243 & 0.955 &  0.407$\pm$0.881 & 3.380 & -1.180$\pm$1.080 & 1.147 &  1.300$\pm$0.935 &  $\:$0.145 \\
 J024008-230915 & 2.23 & 1.635 & 0.767 &  0.608$\pm$1.230 & 1.451 &  0.961$\pm$1.120 & 1.184 &  0.860$\pm$1.230 & -0.180 \\
 J024008-230915 & 2.23 & 1.637 & 0.691 & -0.237$\pm$1.180 & 1.985 &  0.496$\pm$0.772 & 0.688 & -0.812$\pm$1.220 & -0.180 \\
 J024008-230915 & 2.23 & 1.657 & 0.771 & -0.958$\pm$1.300 & 1.468 &  0.766$\pm$0.555 & 0.834 & -0.970$\pm$1.150 & -0.180 \\
 J000344-232355 & 2.28 & 0.452 & 0.883 & -0.706$\pm$0.713 & 2.883 & -1.500$\pm$0.574 & 1.093 & -0.803$\pm$0.690 &  $\:$0.071 \\
 J000344-232355 & 2.28 & 2.185*&   N/A &              N/A$\:\:\:\:\:\:\:\:$ & 0.832 &  7.090$\pm$1.990 &   N/A &    N/A$\:\:\:\:\:\:\:\:$ &  $\:$0.071 \\
 J000344-232355 & 2.28 & 2.187*&   N/A &              N/A$\:\:\:\:\:\:\:\:$ & 1.139 & -0.373$\pm$1.120 &   N/A &              N/A$\:\:\:\:\:\:\:\:$ &  $\:$0.071 \\
 J011143-350300 & 2.40 & 1.182 & 0.806 &  0.298$\pm$0.965 & 0.792 &  0.217$\pm$0.974 & 0.696 &  0.048$\pm$0.925 &  $\:$0.124 \\
 J011143-350300 & 2.40 & 1.350 & 0.856 &  0.175$\pm$0.378 & 1.444 & -2.030$\pm$0.959 & 1.020 & -1.770$\pm$0.974 &  $\:$0.124 \\
 J222006-280323 & 2.41 & 0.942 & 0.947 &  0.686$\pm$1.130 & 2.480 & -1.400$\pm$0.508 & 1.009 &  0.068$\pm$1.010 &  $\:$0.324 \\
 J222006-280323 & 2.41 & 1.556 & 1.059 &  1.050$\pm$0.520 & 2.132 &  1.260$\pm$0.508 & 1.078 &  1.760$\pm$0.666 &  $\:$0.324 \\
 J135038-251216 & 1.44 & 1.439 & 0.759 & -0.071$\pm$0.404 & 3.452 & -0.941$\pm$0.613 & 1.029 & -0.607$\pm$0.297 &  $\:$0.444 \\
 J045523-421617 & 2.66 & 0.908 & 0.896 & -1.240$\pm$0.540 & 5.577 & -0.294$\pm$0.350 & 1.090 & -1.330$\pm$0.553 &  $\:$0.044 \\
 J045523-421617 & 2.66 & 1.858 & 0.627 &  0.069$\pm$3.400 & 6.603 &  0.905$\pm$0.624 & 0.834 &  0.077$\pm$3.540 &  $\:$0.044 \\
 J000448-415728 & 2.76 & 1.542 & 0.759 & -1.150$\pm$1.290 & 1.839 & -4.400$\pm$0.753 & 0.560 & -2.450$\pm$3.750 &  $\:$0.316 \\
 J000448-415728 & 2.76 & 2.168 & 0.709 &  1.640$\pm$1.350 & 3.181 &  0.477$\pm$0.416 & 1.294 &  1.580$\pm$1.380 &  $\:$0.316 \\
 J000448-415728 & 2.76 & 2.302 & 0.895 &  1.600$\pm$0.852 & 6.500 & -0.356$\pm$0.621 & 1.352 & -0.241$\pm$0.522 &  $\:$0.316 \\
 J212912-153841 & 3.29 & 2.022*&   N/A &              N/A$\:\:\:\:\:\:\:\:$ & 0.941 & -1.490$\pm$1.260 &   N/A &              N/A$\:\:\:\:\:\:\:\:$ &  $\:$0.257 \\
\hline
 $(\chi_{v}^{2})_{mean}$ &  &  & \multicolumn{1}{c}{0.82} &  & \multicolumn{1}{c}{2.39} &  & \multicolumn{1}{c}{1.03} & \\

\hline
\end{tabular}
\end{minipage}
\end{table*}

\subsection{Transitions used in a fit}
\label{sec:Transitions used in a fit}

For our best fits we attempted to fit all transitions that show absorption features, if possible. 
Occasionally ionic transitions part of absorption systems  blend with features of
interest in absorption systems from different redshifts. If this contamination is severe, or if there is
evidence for strong contamination by incompletely removed
cosmic rays or telluric features, we remove those transitions from the analysis.
Where this contamination is not severe we model the blended profiles with 
additional Voigt profile components. 

In two cases we were able to
identify the redshift, atomic species and wavelength of the 
blending (i.e. contaminating) transition. 
Absorption from Fe\,\textsc{ii} $\lambda$2382 in an absorption system at z=1.542 fell
close enough to absorption caused by Mg\,\textsc{ii} $\lambda$2803 in the absorption
system at z=0.989 in quasar J000448-415728. Also absorption from Mg\,\textsc{ii}
$\lambda$2803 in an absorption system at z=1.350 fell close enough to absorption 
caused by Mg\,\textsc{ii} $\lambda$2796 in the absorption system at z=1.343 in
quasar J011143-350300.  In these cases we 
fitted both absorption systems simultaneously, increasing the information available to constrain both models. The simultaneous fitting was only required for the absorption systems in our best fits, as the contaminating features fell outside the spectral fitting regions of Variant II. 

The actual transitions used in fitting each absorption system are listed in Table \ref{tab:transitions}.

\subsection{Thermal vs. turbulent fits}
\label{sec:thermturb}

The line-widths for different species can be related turbulently or 
thermally. Previous analyses using the MM method 
have initially constructed Voigt profile fits to absorption systems using a wholly turbulent
broadening mechanism (i.e. $b_{thermal}=0$).
Once the fit of an absorption system was complete, it was re-run using
a wholly thermal broadening mechanism (i.e. $b_{turbulent}=0$). As
long as the values of $\Delta\alpha/\alpha$ did not differ by more
than 1$\sigma$, the fit with the lowest $\chi^{2}$ was chosen as
the final fit \citep{WFC+99,MFW+04}.

\citet{Kin12} followed the same procedure in deriving first turbulent and then thermal 
fits, but then chose to implement a method-of-moments estimator that takes into account
relative differences in the AICc of each fit, and the agreement, or
otherwise, between the values of $\Delta\alpha/\alpha$.

We adopted a method of initially fitting all absorption systems as wholly turbulent. Later in the fitting process, when other ionic species were included in the fit, the line-broadening parameter was switched to wholly thermal in some cases in order to obtain a better model to the absorption system data. If and only if the AICc was reduced in the iteration after switching broadening mechanisms would the thermal broadening mechanism be kept in further fitting iterations. For the 20 cases considered in our final fits, 15 were turbulent for our best fits and Variant II. All fits for Variant I were made using a wholly turbulent broadening mechanism (see Section \ref{sec:Variant I: Underfitting with restricted fitting regions}).

As a consistency check, we also re-analysed all final fits using both broadening mechanisms as in \citet{WFC+99,WMF+01,MFW+04}. None of the fits re-analysed with the alternative broadening mechanism were found to have a lower AICc or $\chi^{2}$ than the original broadening mechanism chosen.

\subsection{General fitting procedure}
\label{sec:General fitting procedure}

We use \textsc{vpfit}, a non-linear $\chi^{2}$ minimization procedure,
which requires the user to supply initial model parameters.
We adopt the following systematic approach, as follows:

\begin{enumerate}
\item The initial velocity structure is built up as a first guess using the
strongest unsaturated transition. \textsc{vpfit }then iterates to minimize
$\chi^{2}$. This initial step is carried
out  with wholly turbulent line-broadening parameters. Additional components are added to improve the
fit iteratively and Voigt profile components are only added if they reduce the AICc. We inspect the value of $\chi^{2}$,
adding components such that $\chi_{\nu}^{2}\approx1$. 
\item Once a satisfactory fit to the first transition, by itself, is obtained, 
that velocity structure is applied to other transitions of the same species.
For example, if the Mg\,\textsc{ii} 
$\lambda$2796 transition
was the first one fitted, Mg\,\textsc{ii} $\lambda$2803
is now fitted simultaneously with Mg\,\textsc{ii} $\lambda$2796. The addition of components is repeated, as in (i).
\item We add transitions of different species to the fit,
tieing physically related parameters. For example,
if Mg\,\textsc{ii} $\lambda$2796 and Mg\,\textsc{ii}
$\lambda$2803 have been fitted, we now add all available
Fe\,\textsc{ii} transitions tieing relevant parameters to the Mg\,\textsc{ii}
transitions.
The exception to this is the Al\,\textsc{iii} doublet. Where available, Al\,\textsc{iii} is 
included in a simultaneous fit but without
any tied N, b or z parameters, such that it contributes independently to 
estimates of $\Delta\alpha/\alpha$.  The reason for not tieing N, b, z is that its higher
ionization potential may result in velocity segregation from the 
other species. The reason
for inclusion is that the difference in \textit{q}-coefficients of the Al\,\textsc{iii} doublet (Table \ref{tab:atomdata}) are large (and different) enough to provide a useful additional sensitivity to a $\Delta\alpha/\alpha$ measurement.
\item In practice three things were taken into consideration in deciding upon
the final model: $\chi_{\nu}^{2}$, the normalised residuals, and the AICc.
For all final models, components are only added if they provide a reduction 
in the numerical value of the AICc, no obvious correlations remain in the
normalised residuals, and $\chi_{\nu}^{2}$ was close to unity.
\item After a final best fit is attained we once again switch the line-broadening mechanism and inspect the AICc. In all cases switching the broadening mechanism after a final fit was attained resulted in an increased the AICc, therefore we kept the previous line-broadening mechanism for each fit.
\end{enumerate}

The entire fitting process is carried out with $\Delta\alpha/\alpha$
set to zero. The introduction of $\Delta\alpha/\alpha$ as a free
parameter is only done once the velocity profile/model to the absorption
system has been finalised.
Once a complete fit is attained \textsc{vpfit }iterates one last time letting $\alpha$
vary as an additional free parameter.

An example fit is shown in the first column of Figure \ref{m1m2m3}.

\subsection{Systems rejected from analysis}
\label{sec:Systems rejected from analysis}

\subsubsection{Absorption system at $z_{abs}=2.185-2.187$ along the line of sight to quasar J000344-232355}

We attempted to fit the absorption complex from $z_{abs}=2.185$ to $z_{abs}=2.187$ as a single absorption system because of the proximity of the absorption features present at each redshift. The absorption features of Fe\,\textsc{ii} $\lambda$2344 at $\lambda$7471$\textrm{\AA}$ and Fe\,\textsc{ii} $\lambda$1608.45 at $\lambda$5126.5$\textrm{\AA}$ both exhibited a shift with respect to the corresponding absorption features present in the other transitions for this absorption system. We found no evidence for any contamination by identified or unidentified interlopers and the addition of components did not resolve the issue. We could not
get a good model to the data and did not arrive at a final best fit. Since a best 
fit, with $\chi_{\nu}^{2}\approx1$ and statistically acceptable normalised 
residuals, could not be attained, this system failed our consistency check 
and was rejected from our analysis.

\subsubsection{Absorption system at $z_{abs}=2.022$ along the line of sight to
quasar J212912-153841}

This system showed absorption by a Mg\,\textsc{ii} doublet, Si\,\textsc{ii} and
multiple Fe\,\textsc{ii} transitions. To provide an accurate measurement of $\Delta\alpha/\alpha$ 
one needs a set of transitions, that when analysed in combination, are sensitive to $\alpha$ variation. Fe\,\textsc{ii} transitions have large, positive \textit{q}-coefficients and Mg\,\textsc{ii}/Si\,\textsc{ii} have small \textit{q}-coefficients (also known as {}``anchors''). The only
anchor lines available in this absorption system are the Mg\,\textsc{ii} doublet and a Si\,\textsc{ii} $\lambda$1526
line. The Mg\,\textsc{ii} doublet falls in a part of the
spectrum heavily contaminated by telluric lines at $\lambda\approx8470\textrm{\AA}$.
Si\,\textsc{ii} $\lambda$1526 falls within the
Lyman-$\alpha$ forest, contaminated by intervening H\,\textsc{i}.
Fe\,\textsc{ii} $\lambda$1608, which could potentially
be used with the other Fe\,\textsc{ii} transitions to constrain $\Delta\alpha/\alpha$ because of it's large and negative \textit{q}-coefficient,
also falls within the forest. The only lines that are free of contamination are the
remaining Fe\,\textsc{ii} transitions, all of which have similar \textit{q}-coefficients. We have therefore rejected this absorption system from our analysis.

\subsection{Best fit results}
\label{sec:Robust error estimates}

Results of individual measurements of $\Delta\alpha/\alpha$, for
each model, are provided in Table \ref{tab:results}.

The weighted mean, using statistical errors only, for our best fits is $\Delta\alpha/\alpha=\left(0.165\pm0.174\right)\times10^{-5}$,
with $\chi_{\nu}^{2}=1.10$. We apply the LTS method to the 20 absorption systems
in our best fit model and find no outliers.

\begin{figure}
 \vspace{0.5cm}
\includegraphics[bb=0 -1 652 467,scale=0.367]{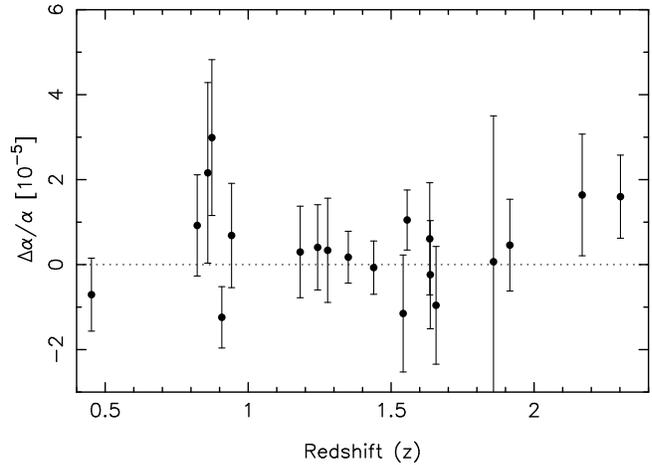}
  \caption{Results from our best fits after increasing error bars
on individual measurements of $\Delta\alpha/\alpha$. $1\sigma$ errors
shown are $\sigma_{total}^{2}=\sigma_{stat}^{2}+\sigma_{rand}^{2}$. 
The weighted mean for this analysis is 
$\Delta\alpha/\alpha=\left(0.223\pm0.226\right)\times10^{-5}$.
\label{m1rsmall}}
\end{figure}

For our main result we
find $\sigma_{rand}=0.479\times10^{-5}$, a weighted mean and error on weighted mean of
\begin{equation}
\Delta\alpha/\alpha=\left(0.223\pm0.226\right)\times10^{-5},
\end{equation}

\noindent with a $\chi_{\nu}^{2}$ of 1.10. No outliers were found. Individual values of $\Delta\alpha/\alpha$
and robust $1\sigma$ error estimates $(\sigma_{total})$ are presented
in Figure \ref{m1rsmall}. The results of our best fitting model, using purely statistical errors $(\sigma_{stat})$, are presented in Figure \ref{m1vm2raw} as black circles.

\subsubsection{Comparison with \citet{CSP+04} and \citet{SCP+07}}

The absorption systems in our best fit model have been previously analysed by C04. They reported $\Delta\alpha/\alpha= -0.06 \pm 0.06 \times 10^{-5}$. Following a series of criticisms \citep{MWF07,MWF08}, \citet{SCP+07} revised
the previous C04 result to $\Delta\alpha/\alpha= 0.01 \pm 0.15
\times 10^{-5}$, although not all criticisms were addressed in the revised analysis.

Our best fit results do not conflict with the mean $\Delta\alpha/\alpha$ value of the updated C04 analysis reported in \citet{SCP+07}. While our errors on the weighted mean are larger, and include a random systematic term, there is no conflict in the results. However, our analysis of model Variants I $\&$ II shows that there are significant systematic effects which remain unaccounted for in the C04 and \citet{SCP+07} results (see Section \ref{sec:Comparison of different fitting Methodologies}).

\subsubsection{Comparison with the King et al. (2012) dipole variation}

In order to compare the results from our best fits with the
possible dipole model reported by \cite{Kin12}, we proceed as follows.
We calculate 
\begin{equation}
\chi^{2}=\sum_{i=1}^{N}\frac{(x_{i}-y_{i})^{2}}{\sigma_{i}^{2}},
\end{equation}

\noindent where $x_{i}$ is the $i$th measurement of $\Delta\alpha/\alpha$ for our best fits, 
$y_{i}$ is the corresponding value of $\Delta\alpha/\alpha$ predicted for that position 
in the sky according to the dipole model parameters found in \citet{Kin12} and $\sigma_{i}$ is the 
1$\sigma$ statistical uncertainty associated with that measurement from our best fit results. For the dipole model of $\Delta\alpha/\alpha=Acos(\Theta)+m$ we use the parameters reported in \citet{Kin12} of angular amplitude $A=0.97_{-0.20}^{+0.22}\times10^{-5}$,
pointing in the direction $RA=(17.3\pm1.0)$hr, $dec=(-61\pm10)^{\circ}$
and having a monopole term of $m=(-0.18\pm0.08)\times10^{-5}$.

Following \citet{Kin12} we also use the LTS method 
to calculate $\sigma_{rand}$ for our best fit results with respect to the dipole model and use this value as a measure of 
scatter with respect to the dipole model. 

We find a $\chi_{\nu}^{2}=0.99$ for our best fits with respect to the dipole model. Using the 
LTS method we calculate $\sigma_{rand}=0.434\times10^{-5}$. Comparing these values to those 
calculated with respect to a weighted mean, $\chi_{\nu}^{2}=1.10$ and $\sigma_{rand}=0.479\times10^{-5}$, 
we find that that the data do not conflict with the dipole results reported in \citet{Kin12}.

To make a direct comparison with the weighted mean calculations in Table \ref{table:weighted means}, we adopt the dipole parameters above and calculate the expected values of $\Delta\alpha/\alpha$ for each data point in the best fit sample. For these expected dipole values of $\Delta\alpha/\alpha$ we take the errors from our best fits, that is, the statistical errors on each point returned by \textsc{vpfit}. The expected weighted mean for this sample is $(\Delta\alpha/\alpha)_{I}=\left(0.205\pm0.174\right)\times10^{-5}$. These results agree with the statistical analysis of \citet{Kra+14} who found general consistency between the 
results of \citet{MWF03}, \citet{Kin12} and \citet{SCP+07}.

\section{Comparison of different fitting Methodologies}
\label{sec:Comparison of different fitting Methodologies}

We employ three different fitting methodologies,
which we name our best fits, Variant I and Variant II.
By modelling the absorption systems in different ways we are able to
compare modelling techniques and their effect on the measurement of the variation of
$\alpha$ and associated errors. Our best fits have been discussed previously. Variant I employs the fitting
approach of C04 and Variant II mimics a very
specific aspect of the C04 Voigt profile fitting methodology. 
Details of each model, and common methods, are provided in the following subsections.

\subsection{Variant I: Under-fitting with restricted fitting regions}
\label{sec:Variant I: Underfitting with restricted fitting regions}

In this model we emulate the Voigt profile fitting procedure
used by C04. \citet{MWF08} have demonstrated that absorption systems that are 
fit with too few Voigt profile components, as is the case for the C04 analysis, will produce spurious results of 
$\Delta\alpha/\alpha$. To examine if these fitting choices have an effect on the measurement of $\Delta\alpha/\alpha$ and errors we have used the model parameters, transitions and fitting regions listed in 
C04 as a basis for Variant I. That is, we have used $z$ and $b$ components 
provided in their analysis
as starting guesses. The $b$-parameters, $z$'s and transitions
used in a fit are provided in C04. However the column
densities and fitting ranges are not provided directly in C04.
Column density starting guesses were chosen by fixing all parameters
in an absorption system except column densities, and letting \textsc{vpfit
}iteratively find a $\chi^{2}$ minimum for the column densities alone.
All fits for Variant I use the turbulent line broadening mechanism. 
We have used the same spectra, extractions, error arrays
and atomic data as were used for our best fit models. We employ \textsc{vpfit
}as our $\chi^{2}$ minimization algorithm.

\begin{figure}
 \vspace{0.5cm}
\includegraphics[bb=0 -1 656 467,scale=0.367]{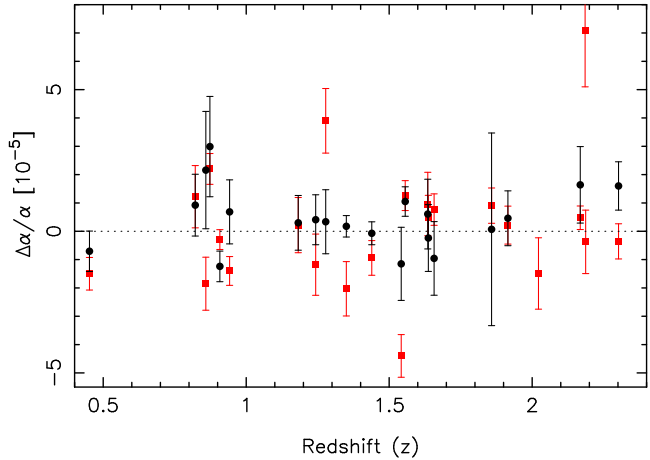}
  \caption{Results from our best fitting model (black
circles) are compared with results of Variant I (red squares). All
errors shown are $1\sigma$ and purely statistical $(\sigma_{stat})$.
Note the increased scatter of Variant I values of $\Delta\alpha/\alpha$,
as is clear from the plot and the values of $\chi_{\nu}^{2}$ for
each fit (see Figure \ref{chi}). For our best fits, only taking into account raw statistical errors,
$\chi_{\nu}^{2}=1.10$ for the weighted mean. For Variant I, only taking into account raw
statistical errors, $\chi_{\nu}^{2}=5.11$ and the weighted mean is $\Delta\alpha/\alpha=\left(-0.047\pm0.138\right)\times10^{-5}$.
\label{m1vm2raw}}
\end{figure}

This model differs from our best fits in many important respects:

\begin{enumerate}
\item Wavelength ranges are restricted so as to only include the strongest
parts of absorption, sometimes so much so that little or no continuum
regions are included. This means that parts of the absorption
system that affect the model are neglected (for example, see Figure \ref{m1m2m3}, 
the feature at -35$\>$km$\;$s\textsuperscript{-1} in Mg\,\textsc{ii} $\lambda$2796 and 
Mg\,\textsc{ii} $\lambda$2803 in Variant I). Specifically, spectral fitting regions were chosen to match those present in the plots provided in C04.
\item In many cases, our best fits make use of ionic transitions that are 
excluded from Variant I.  These are generally transitions that show weak absorption.  An example of
this is illustrated in Fig. 1, where we include an Mg\,\textsc{i} transition in our best Voigt profile fit.
The statistical impact of including all possible transitions is 
small but generally helps to improve the accuracy of the $\Delta\alpha/\alpha$ measurement, as we 
demonstrate in Section \ref{sec:Quantifying the effect of reduced spectral fitting regions}.
\item Table \ref{tab:transitions} shows that 27 of the transitions included
in Variant I fits are not included in our best fits. In
all excluded cases, this is because close inspection reveals problems
such as blending with atmospheric absorption features or contamination by unidentified absorption that could not be adequately modelled.  The inclusion of
contaminated transitions has the potential to produce spurious results
of $\Delta\alpha/\alpha$. An example of this is the absorption system at $z_{abs}=1.556$ towards quasar J222006-280323, Si\,\textsc{ii} $\lambda$1526 absorption is blended with Lyman forest absorption and Mg\,\textsc{ii} $\lambda$2803 absorption is blended with the $O_{2}$ band.
\item The vast majority of fits in Variant I contain far fewer parameters
than the fits for our best model, i.e. Variant I is under-fit, as has been demonstrated 
in \citet{MWF08}. This also has
potential to produce spurious results of $\Delta\alpha/\alpha$.
\end{enumerate}

\begin{figure}
 \vspace{0.5cm}
\includegraphics[bb=0 -1 652 467,scale=0.367]{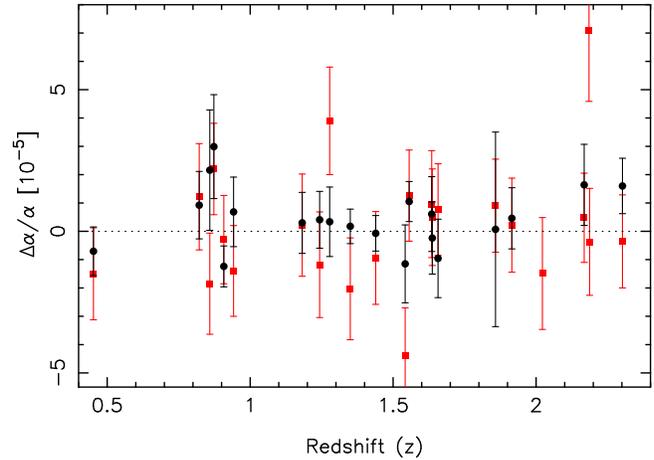}
  \caption{Comparison between our best fitting model (black circles) and Variant I (red squares) after increasing error bars
on individual results of $\Delta\alpha/\alpha$. $1\sigma$ errors
shown are $\sigma_{total}^{2}=\sigma_{stat}^{2}+\sigma_{rand}^{2}$. The weighted mean for Variant I using statistical and systematic errors is $\Delta\alpha/\alpha=\left(0.002\pm0.361\right)\times10^{-5}$ with $\chi_{\nu}^{2}=1.22$.
\label{m1vm2robust}}
\end{figure}

An example fit of Variant I is shown in column two of Figure \ref{m1m2m3}.

\subsection{Variant II: Restricted spectral fitting regions}
\label{sec:Variant II: Restricted spectral fitting regions}

We have opted to include a third modelling technique to investigate
the effects of a specific aspect of fitting present in C04.
Variant II is the same as our best fits in all respects except
one; the spectral fitting regions are reduced to exactly match those of C04. 
Modelling in this manner allows us to quantitatively investigate
the effects of reduced spectral fitting regions on $\Delta\alpha/\alpha$
and associated errors by a comparison of our best fits with Variant II. We
use the same Voigt profile fits, spectra, error arrays
and atomic data as are used for our best fits. The only change is the spectral
fitting regions. Where a Voigt profile component does not fall within
the chosen fitting regions, we do not use that component in our fit.
Spectral fitting regions are adjusted and components falling
outside those regions dropped. \textsc{vpfit} iterates a new minimum
of $\chi^{2}$ and a new value of $\Delta\alpha/\alpha$ and associated
error.

An example of a Variant II absorption system is shown in column three of Figure \ref{m1m2m3}.

\subsection{Results for Variants I and II}
\label{sec:Results for Models II and III}

\subsubsection{Variant I: Under-fitting with restricted fitting regions}
\label{sec:Results Variant I: Underfitting with restricted fitting regions}

The weighted mean, using statistical errors only, for Variant I is $\Delta\alpha/\alpha=\left(-0.047\pm0.138\right)\times10^{-5}$,
with $\chi_{\nu}^{2}=5.11$. We apply the LTS method to the 23 absorption systems
in Variant I and find no outliers. These data points are presented in 
Figure \ref{m1vm2raw} as red/grey squares.

Allowing for non-zero $\sigma_{rand}$, we
find $\sigma_{rand}=1.519\times10^{-5}$, a weighted mean and error on weighted mean of
\begin{equation}
\Delta\alpha/\alpha=\left(0.002\pm0.361\right)\times10^{-5},
\end{equation}

\noindent with a $\chi_{\nu}^{2}$ of 1.22. No outliers were found. Individual values of $\Delta\alpha/\alpha$
and robust $1\sigma$ error estimates $(\sigma_{total})$ are presented
in Figure \ref{m1vm2robust} as red/grey squares.

\subsubsection{Variant II: Reduced spectral fitting regions}
\label{sec:Variant II: Reduced spectral fitting regions}

\begin{figure}
 \vspace{0.5cm}
\includegraphics[bb=0 -1 652 467,scale=0.367]{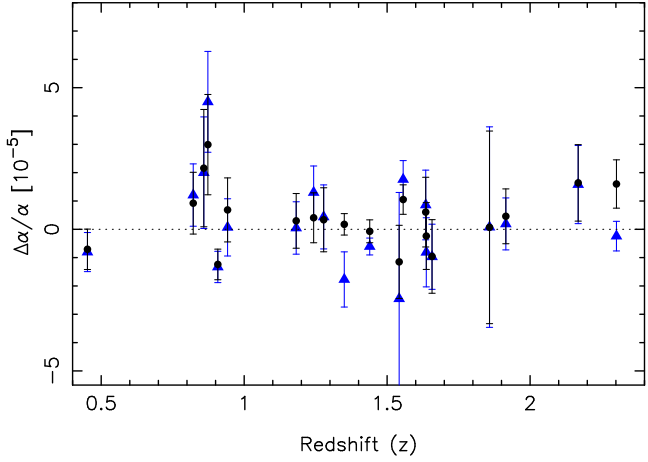}
  \caption{Comparison between our best fit results (black circles) and Variant II results (blue triangles). All errors shown are $1\sigma$ and purely statistical $(\sigma_{stat})$,
derived solely from the relevant covariance matrix diagonal terms
calculated by \textsc{vpfit}. For Variant II, only taking into account raw
statistical errors, $\chi_{\nu}^{2}=1.83$ and the weighted mean is $\Delta\alpha/\alpha=\left(-0.200\pm0.174\right)\times10^{-5}$.
While Variant II does show an increase in the scatter of individual
values of $\Delta\alpha/\alpha$, both models are in good agreement with each other.
\label{m1vm3raw}}
\end{figure}

The weighted mean, using statistical errors only, for Variant II is $\Delta\alpha/\alpha=\left(-0.200\pm0.174\right)\times10^{-5}$,
with $\chi_{\nu}^{2}=1.83$. We apply the LTS method to the 20 absorption systems
in Variant II and find no outliers. These data points are presented in 
Figure \ref{m1vm3raw} as blue/grey triangles.

\begin{figure}
 \vspace{0.5cm}
\includegraphics[bb=0 -1 652 467,scale=0.367]{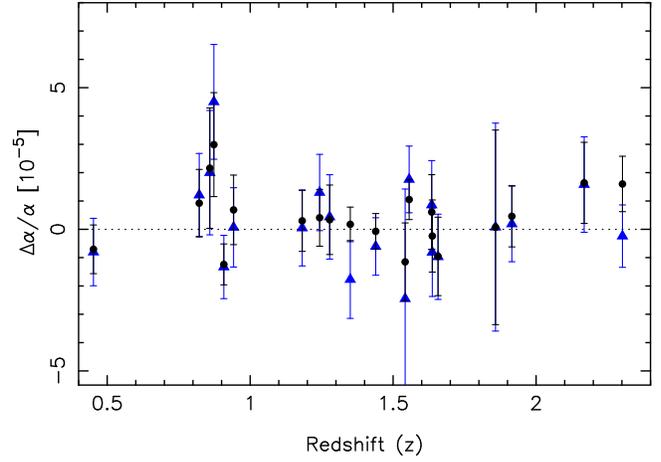}
  \caption{Comparison between our best fit results (black circles) and Variant II results (blue triangles) after increasing error bars
on individual results of $\Delta\alpha/\alpha$. $1\sigma$ errors
shown are $\sigma_{total}^{2}=\sigma_{stat}^{2}+\sigma_{rand}^{2}$. The weighted mean for Variant I using statistical and systematic errors is $\Delta\alpha/\alpha=\left(0.107\pm0.319\right)\times10^{-5}$ with $\chi_{\nu}^{2}=0.79$.
\label{m1vm3robust}}
\end{figure}

Allowing for non-zero $\sigma_{rand}$, we
find $\sigma_{rand}=0.970\times10^{-5}$, a weighted mean and error on weighted mean of
\begin{equation}
\Delta\alpha/\alpha=\left(0.107\pm0.319\right)\times10^{-5},
\end{equation}

\noindent with a $\chi_{\nu}^{2}$ of 0.79. No outliers were found. Individual values of $\Delta\alpha/\alpha$
and robust $1\sigma$ error estimates $(\sigma_{total})$ are presented
in Figure \ref{m1vm3robust} as red/grey squares.

\subsubsection{Quantifying the effect of reduced spectral fitting regions}
\label{sec:Quantifying the effect of reduced spectral fitting regions}

The only differences between the modelling techniques of our best fits and those of Variant II
are the sizes of the spectral fitting regions chosen for the fits. Voigt profile 
components that fall outside these regions are necessarily excluded. The vast majority of these Voigt profile components are weak components falling on the outskirts of the main, stronger features
of the absorption system. In some cases strong absorption features are ignored due to the choice of fitting regions, such as the absorption features at $v$ $\approx$ -50$\>$km$\;$s\textsuperscript{-1} in the middle panel of Figure \ref{m1m2m3}. By comparing the results of our best fits with
those of Variant II we can directly test the effect of reduced spectral
fitting regions on the determination of $\Delta\alpha/\alpha$. Table
\ref{table:weighted means} lists the weighted means with purely statistical errors and
also the weighted means that have accounted for extra scatter by increasing
the $1\sigma$ error bars via $\sigma_{rand}$.

Figure \ref{m1vm3raw} shows a plot of results comparing our best fit results with those of Variant II.
Overall, there is good agreement. $\Delta\alpha/\alpha$ values generally
agree very well between both models, although there is more scatter for the Variant II points.

We note that our best fit absorption systems at $z_{abs}=1.542$
and $z_{abs}=1.350$ are contaminated by intervening absorption systems that reside in different redshifts, and happen to fall on that part of the spectrum 
(these absorption systems and their parameter values can be viewed supplementary 
material). The absorption system towards J000448-415728 at $z_{abs}=1.542$ 
was fitted simultaneously with another at $z_{abs}=1.989$ because the 
transition Mg\,\textsc{ii} $\lambda$2803 in the latter blends with 
Fe\,\textsc{ii} $\lambda$2383 in the former. Also the absorption system 
towards  at $z_{abs}=1.350$ was fitted simultaneously with a magnesium 
doublet at $z_{abs}=1.344$ because the Mg\,\textsc{ii} $\lambda$2803 
transition in the latter blends with Mg\,\textsc{ii} $\lambda$2796 in the 
former. Systems $z_{abs}=1.542$ and $z_{abs}=1.350$ are $not$ treated as contaminated 
in Variant II, as the wavelength fitting ranges are restricted enough so 
as to exclude the contaminating absorption features. Therefore it is inappropriate 
to compare the results of our best to the Variant II results for absorption systems $z_{abs}=1.542$
and $z_{abs}=1.350$.

\subsubsection{Extra scatter when fitting regions are restricted}
\label{sec:Extra scatter when fitting regions are restricted}

As previously discussed, 
we calculate the error on each
$\Delta\alpha/\alpha$ measurement as a quadrature addition of two terms (Section \ref{sec:The Least Trimmed Squares (LTS) method}), one 
derived from the covariance matrix at the best fit, $\sigma_{stat}$, the other 
allowing for additional sources of error such as velocity structure ambiguities, 
calibration and other uncertainties, $\sigma_{rand}$. Since our best fit model and Variant II are identical in fitting methodology apart from a variation of fitting regions, the comparison between the two models can be 
parametrized in terms of $\sigma_{rand}$. If restricting the fitting region 
has the effect of increasing the scatter on $\Delta\alpha/\alpha$ measurements this would be indicated as an associated increase in $\sigma_{rand}$.

Because of the two systems removed, as discussed in the previous section, there are 18 
points common to both samples. The result for our best fits is
\begin{equation}
(\sigma_{rand})_{Best}=0.457\times10^{-5}
\end{equation}

\noindent for 18 absorption systems. The result for Variant II is
\begin{equation}
(\sigma_{rand})_{II}=0.867\times10^{-5}
\end{equation}

\noindent for 18 absorption systems. We see an $\approx90\%$ increase in $\sigma_{rand}$
for Variant II over our best fits. Since the only difference between the two models is
the choice of fitting regions, this extra scatter exhibited by
Variant II is a direct consequence of restricting spectral fitting regions. This result illustrates 
the necessity of careful choice of fitting regions and the significance to which poor choice of fitting 
regions can lead to spurious measurements of $\Delta\alpha/\alpha$.

The increase in $\sigma_{rand}$ due to narrow fitting regions is perhaps at first sight surprising. 
However, Figure 1 illustrates the reason. It can be seen that for Variant II, the relatively strong 
absorption seen in Mg\,\textsc{ii} for velocities less than about -25$\>$km$\;$s\textsuperscript{-1} 
must impact on the stronger components at velocities immediately to the right of -25$\>$km$\;$s\textsuperscript{-1}.
In turn, modifications in the range -25$\>$km$\;$s\textsuperscript{-1} < $v$ < 0$\>$km$\;$s\textsuperscript{-1} 
will have a knock-on effect further into the complex, and so on.
Including the full fitting range allows \textsc{vpfit} to properly evaluate redshift parameter 
errors. However, if a narrow fitting range is used, \textsc{vpfit} is ``tricked'' into computing 
artificially small redshift errors. Figure \ref{m1vm3raw} and \ref{m1vm3robust} illustrate further 
the consequence of inappropriately narrow fitting regions. The blue triangles (Variant II) exhibit greater scatter than the black points.

\subsubsection{Quantifying the effect of excluding weak transitions from a fit}
\label{sec:Quantifying the effect of excluding weak transitions from a fit}

The measurement of $\Delta\alpha/\alpha$ is of course most sensitive to strong, unsaturated, absorption features. We have used transitions that show relatively weak absorption features. This is justified by the fact that the statistical uncertainty of the fits using these transitions is improved, albeit marginally.

For example, Variant I does not include three weakly
absorbing transitions (Mg\,\textsc{i} $\lambda$2852
and Fe\,\textsc{ii} $\lambda$2344/2587) in the fit
of the absorption system at $z_{abs}=0.942$ along the line of sight
to quasar J222006-280323 (see Variant I in Figure \ref{m1m2m3}). In our best fits to the
data we include these transitions (see our best fit in Figure
\ref{m1m2m3}) and find $\Delta\alpha/\alpha=(0.686\pm1.130)\times10^{-5}$
for this system. Removing Mg\,\textsc{i} $\lambda$2852
and Fe\,\textsc{ii} $\lambda$2344/2587 from the fit
and re-calculating $\Delta\alpha/\alpha$ we find $\Delta\alpha/\alpha=(0.627\pm1.390)\times10^{-5}$.
There is an increase in the uncertainty of $\Delta\alpha/\alpha$ when these weak transitions are removed
and the value of $\Delta\alpha/\alpha$ does not significantly change.

Another example is the absorber at $z_{abs}=1.637$ along the line of sight to quasar J024008-230915. For our best fit of this absorber $\Delta\alpha/\alpha=(-0.237\pm1.180)\times10^{-5}$ . Variant I excludes the weakest transition in this absorption system, Mg\,\textsc{i} $\lambda$2852. Removing this transition from our best fit and re-calculating $\Delta\alpha/\alpha$ we find $\Delta\alpha/\alpha=(-0.253\pm1.200)\times10^{-5}$.

\begin{figure}
\includegraphics[bb=0 140 656 467,scale=0.367]{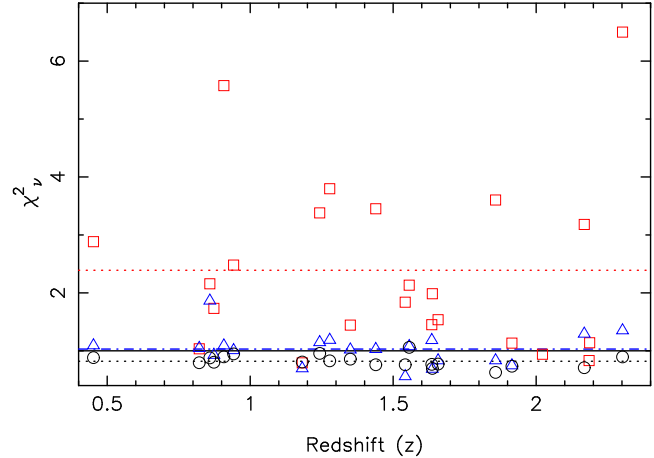}
 \vspace{1.5cm}
  \caption{$\chi_{\nu}^{2}$ comparison. Black circles show individual values
   for our best fit results, red squares show individual values for Variant I and blue triangles show individual
values for Variant II. Black dotted line, red dotted line and blue dash-dotted line 
show the means of $(\chi_{\nu}^{2})_{Best}=0.82$, $(\chi_{\nu}^{2})_{I}=2.39$
and $(\chi_{\nu}^{2})_{II}=1.03$, for our best fits, Variants I and II respectively.
Our best fits and Variant II show very few points in excess of unity, suggesting that the 
absorption systems are statistically acceptable fits for these models. 
Variant I shows many points with values of $\chi_{\nu}^{2}$ well in excess of 
unity, suggesting that these models are not statistically good fits to the data.
\label{chi}}
\end{figure}

The two examples above show that, for these particular cases,
when fitting multiple transitions in the same species, 
including the weakest transitions or transition causes
the 1$\sigma$ error estimate on $\Delta\alpha/\alpha$ to reduce by about $10\%$.
The actual value of $\Delta\alpha/\alpha$ shifts by a negligible 
amount compared to the 1$\sigma$ error estimate.  We have not attempted
to generalise this further, although one would expect
it to be beneficial to include \textit{all} transitions of all strengths for a given species
in cases like this, where stronger lines in that species are clear detections.

\begin{table*}
 \begin{minipage}{172mm}
  \caption{Summary and comparison of results for the 23 absorption systems studied
in this work. Column 1 enumerates the different results. Column 2
gives descriptions of the sample. Column 3 gives the number of absorption
systems analysed in each sample. Column 4 gives $\chi_{\nu}^{2}$
for each sample. Column 5 gives the weighted mean and $1\sigma$ uncertainty
of each sample, in units of $10^{-5}$. Column 6 shows the raw statistical error 
on the weighted mean and column 7 shows $\sigma_{rand}$ as calculated by the LTS method. 
Rows 1-3 only show
statistical $1\sigma$ uncertainties $(\sigma_{stat})$. Rows 4-6
show robust error estimates, estimated by increasing uncertainties
by adding a constant in quadrature to match the scatter $(\sigma_{total})$.}
  \begin{tabular}{@{}llllrrr@{}}
  \hline
 I & Sample & $\textit{N}_{abs}$ & $\chi_{v}^{2}$ & $\Delta\alpha/\alpha[10^{-5}]$
                          & $\sigma_{stat}[10^{-5}]$ & $\sigma_{rand}[10^{-5}]$ \\
  \hline
1 & Best fits: Optimal fits, statistical errors                      & 20 & 1.10   &  0.165$\pm$0.174 & 0.174 & -\tabularnewline
2 & Variant I: C04 emulations, statistical errors                & 23 & 5.11       & -0.047$\pm$0.138 & 0.138 & -\tabularnewline
3 & Variant II: Restricted fitting regions, statistical errors   & 20 & 1.83       & -0.200$\pm$0.174 & 0.174 & -\tabularnewline
4 & Best fits - Robust errors                                     & 20 & 0.77$^a$   &  0.223$\pm$0.226 & 0.174 & 0.479\tabularnewline
5 & Variant I - Robust errors                                    & 23 & 1.22$^a$   &  0.002$\pm$0.361 & 0.138 & 1.519\tabularnewline
6 & Variant II - Robust errors                                   & 20 & 0.79$^a$   &  0.107$\pm$0.319 & 0.174 & 0.970\tabularnewline
 \hline                                  
 \multicolumn{7}{L{168mm}}{$^{a}$Note the expectation values of $\chi^2_{\nu}$ are not unity.
The LTS method we have used produces robust error estimates by 
estimating the additional $\sigma_{rand}$ term from
a fraction ($f=0.85$ in our case) of the entire sample such that the observed 
scatter about the weighted mean matches the theoretical expectation for
that subset.  To illustrate this, the theoretical expectation of
$\chi^2_{\nu}$ for our best fits, when LTS is applied to a subset of the entire sample 
(17 data points in this case), is $0.44$.  $\sigma_{rand}$ is derived by matching the
observed and theoretical $\chi^2_{\nu}$.
The $\sigma_{rand}$ derived in this way is then applied to
the {\it whole} sample.}    \\
 \label{table:weighted means}
  \end{tabular}
 \end{minipage}
\end{table*}

\subsubsection{Quantifying the effect of under-fitting}
\label{sec:Quantifying the effect of underfitting}

The fitting methodology of Variant I differs from that of our best fits in three important respects, \emph{i)} differences
in spectral fitting regions, \emph{ii)} differences in transitions included
in a fit, and \emph{iii)} differences in modelling of velocity structure,
that is, Variant I uses fewer velocity components, even after allowing for the restricted 
spectral fitting regions. Here, by comparing our best fits 
with Variant I, and noting of the effects of points \emph{i)} and \emph{ii)}
above, we show that under-fitting has the most significant (and adverse) impact
on estimating $\Delta\alpha/\alpha$.

Comparing overlapping fitting regions between out best fit model with Variant I,
i.e. comparing like with like, the AICc method results (our best fits) in a $60\%$ increase in the number of components compared to the Variant I model. For example, compare the middle panel (Variant I) and the third 
panel (Variant II) in Figure \ref{m1m2m3}. Variant II contains twice as many velocity components 
as Variant I. The mean $\chi_{\nu}^{2}$ for Variant I is $(\chi_{\nu}^{2})_{mean}=2.39$,
reflecting the fact that Variant I does not, in general, attain statistically
acceptable fits to absorption systems. Figure \ref{chi} shows a comparison
of the values of $\chi_{\nu}^{2}$ for each individual absorption system for
all three models.

\subsubsection{Extra scatter in Variant I}
\label{sec:Extra scatter in Variant I}

We further quantify the effect of under-fitting on the data by comparing
the value of $\sigma_{rand}$ for each model. We follow the same procedure as in Section
\ref{sec:Extra scatter when fitting regions are restricted}. We also only calculate $\sigma_{rand}$ for a subset of systems
that are common to both our best fits and Variant I. The result for our best fits is
\begin{equation}
(\sigma_{rand})_{Best}=0.457\times10^{-5}
\end{equation}

\noindent for the 18 systems that are in common. The result for Variant I is
\begin{equation}
(\sigma_{rand})_{I}=1.300\times10^{-5}.
\end{equation}

\noindent We therefore see from the above, and from Section \ref{sec:Extra scatter when fitting regions are restricted}, that compared to
independent fits, restricting the fitting regions (Variant II) increases
$\sigma_{rand}$ by from 0.457$\times10^{-5}$ to 0.867$\times10^{-5}$, and under-fitting further increases
$\sigma_{rand}$ from 0.867$\times10^{-5}$ to 1.300$\times10^{-5}$, i.e. both fitting choices have a
significant adverse effect on the fitting accuracy.

\subsubsection{Effect of under-fitting on $\sigma_{stat}$ and $\sigma_{rand}$}
\label{sec:Effect of under-fitting on sigmastat and sigmarand}

A well-known feature of non-linear least-squares modelling is that
parameter error estimates derived from the covariance matrix at the
best-fit solution (as is done by \textsc{vpfit}) are only reliable when a number
of conditions are met, one of which is that the model is a
statistically valid representation of the data.  If this is not the
case, one would not expect a non-linear least-squares procedure to
return meaningful parameters and parameter errors.

Inspecting Table \ref{tab:results}, comparing our best fit results with those of Variant I, it can be seen that
in some cases our best fits yield substantially larger error
estimates than Variant I (C04).  However, in other cases, the error estimates
are seen to be in good agreement.  In a few cases, the error estimate
we derive from our best fits are actually smaller than those for Variant I.
The explanations for this are as follows.  Our best fits contain, on average,
a higher number of velocity components than Variant I.  When close blending
occurs in the transitions most sensitive to an alpha variation, i.e.
the stronger, often central components, we clearly expect a substantial
increase in the $\Delta\alpha/\alpha$ error estimate.  On the other hand, when the additional
components fall sufficiently far from the main components determining alpha,
little difference is expected between the best fit and Variant I $\Delta\alpha/\alpha$ error estimates.
Finally, because our best fits include additional transitions not used in Variant
I, {\it and} when the additional velocity components do not fall in the
stronger lines, we expect the best fit $\Delta\alpha/\alpha$ error estimates to reduce compared to
Variant I.

The following example illustrates a substantial result of under-fitting exhibited 
within this dataset. We can isolate the impact of under-fitting the data by 
comparing Variants I and II and use, as an example, the $z_{abs}=0.942$ system 
towards J222006-280323 (shown in Figure \ref{m1m2m3}) to illustrate the problems. 
Figure \ref{m1m2m3} and Table \ref{tab:transitions} show that for this system, our best fits 
included 3 additional weak transitions Mg\,\textsc{i} $\lambda$2852, 
Fe\,\textsc{ii} $\lambda$2586 and Fe\,\textsc{ii} $\lambda$2344.  To directly 
assess the impact of under-fitting alone, and using our best fit as a template, 
we re-fit this system excluding those transitions and adopting the 
restricted wavelength fitting regions of Variant I.  The results for the refit are
$\Delta\alpha/\alpha=(-0.047\pm1.140)\times10^{-5}$.  When including the 3 weak 
transitions, our original Variant II result gave 
$\Delta\alpha/\alpha=(0.068\pm1.010)\times10^{-5}$).  Variant I, which now 
comprises exactly the same transitions and fitting regions gives 
$\Delta\alpha/\alpha=(-1.400\pm0.508)\times10^{-5}$.
The uncertainties quoted here are statistical only, i.e. the errors
returned provided by \textsc{vpfit}.  The Variant I errors are a factor of two
too small and the actual $\Delta\alpha/\alpha$ value for Variant I differs 
significantly from our best fit value.  

The problems caused by under-fitting are highlighted further still when one
considers the scatter in the sample as a whole.  The Variant I sample exhibits 
far more scatter about a mean value than Variant II (red squares in Figure \ref{m1vm3robust}).
The values of $\sigma_{rand}$ for the Variant I and II samples are 1.688$\times10^{-5}$ 
and 0.916$\times10^{-5}$ respectively. Using the example absorption system from the 
previous paragraph, if we now attempt to estimate a random systematic error term we find 
that it has ballooned for Variant I, due to the excess scatter present in the data. The final 
error estimate, including statistical and random systematic terms added in quadrature, for absorption system 
$z_{abs}=0.942$ for Variant I is 1.602$\times10^{-5}$, while it is only 1.227$\times10^{-5}$ 
for our best fit. This example highlights
two important, albeit subtle, considerations.  First, under-fitting
results in meaningless parameter errors.  Second, under-fitting results
in greater scatter overall.

\section{Conclusions}
\label{sec:Conclusions}

Our work investigates profile fitting methodologies and provides a new analysis of 23 quasar absorption systems. Our findings are as follows:

\begin{enumerate}
\item It is important to use a reproducible and systematic approach to estimating
the number of absorbing components fitted to an absorption complex.  Here we have
used the AICc statistic to estimate the number of components required. Our best fits, on average, use $\approx60\%$ more Voigt profile components to adequately model an absorption region of the same size
as Variant I. When an insufficient number of components are used to model an absorption system the numerical value of $\Delta\alpha/\alpha$ changes significantly (i.e. the value of $\sigma_{rand}$ is much smaller for the ensemble of our best fits compared to Variant I - see Section \ref{sec:Quantifying the effect of underfitting}). The larger number of components results in a much smaller scatter amongst an ensemble of $\Delta\alpha/\alpha$ estimates. Also, the increased number of components means more model
parameters which in turn causes the estimated statistical uncertainty on $\Delta\alpha/\alpha$ to increase,
as one would expect.  Conversely, underestimating the number of free parameters in
the model produces a significantly smaller estimated uncertainty on $\Delta\alpha/\alpha$, but if the model is
incorrect, both parameter estimates and associated errors are meaningless. Figure \ref{m1m2m3} gives a strong example of this: for that absorption system our best fit gives a result of $\Delta\alpha/\alpha=\left(0.686\pm1.130\right)\times10^{-5}$ and the Variant I model's results are $\Delta\alpha/\alpha=\left(-1.400\pm0.508\right)\times10^{-5}$.

\item When modelling absorption features, it is important that fitting regions
cover all relevant parameters, including continuum regions without detectable absorption.  Put differently, it is important to avoid 
subjective selection of what may appear to be the ``most interesting'' or 
strongest features.  It is incorrect to assume that weaker absorption features nearby
have an insignificant impact on the best-fit model parameters and their associated errors. Restricting fitting regions in this way (our Variant II model) produces a $90\%$ increase in $\sigma_{rand}$ when compared to the results of our best fit model.

\item Combining the impact of both restricted fitting regions and under-fitting results in an increase of the error
estimate on the weighted mean of the sample of $60\%$ when compared to our best fit model.

\item Using our robust error
estimates, we find a final weighted mean our independent analysis of the \cite{CSP+04} dataset (our best fits) of
$\Delta\alpha/\alpha=\left(0.223\pm0.226\right)\times10^{-5}$. 
The weighted mean of our best fits, quoting statistical errors only, is found 
to be $\Delta\alpha/\alpha=\left(0.165\pm0.174\right)\times10^{-5}$.

\item Our best results
are also consistent with the dipole model of $\alpha$ variation
presented in \cite{Kin12}.

\item The analysis of our best fits to the absorption systems are
consistent with the weighted mean of the much larger sample of 154
absorption systems analysed by \cite{Kin12} of
$\Delta\alpha/\alpha=\left(0.208\pm0.124\right)\times10^{-5}$.
\end{enumerate}

\section*{Acknowledgements}

We are grateful to the referee, Paulo Molaro, for very helpful comments which significantly improved the manuscript. We would like to thank R. F. Carswell  and J. C. Berengut for advice and insightful comments on early version of the manuscript. M. T. Murphy thanks the Australian Research Council for Discovery Project grant DP110100866 which supported this work. M. R. Wilczynska has been supported in part by an Australian Postgraduate Award.

\bibliographystyle{mn2e}
\bibliography{example}

\appendix

\label{lastpage}

\end{document}